\documentclass[12pt]{article}
\usepackage[english]{babel}
\usepackage{latexsym}
\usepackage{amsfonts}
\begin{document}
\begin{center}
{\Large \bf On the long-range gravity in warped backgrounds}\\

\vspace{4mm}

Mikhail N.~Smolyakov\\

\vspace{4mm}

Physics Department, Moscow State University,\\ Vorob'evy Gory,
119992 Moscow, Russia
\\
\end{center}

\begin{abstract}
In this paper the Randall-Sundrum model with brane-localized
curvature terms is considered. Within some range of parameters a
compact extra dimension in this model can be astronomically large.
In this case the model predicts small deviation from Newton's law
at astronomical scales, caused by the massive modes. The existence
of this deviation can result in a slight affection on the
planetary motion trajectories.\vspace{0.3cm}
\end{abstract}

\section{Introduction}
Nowadays models with brane-localized curvature terms are widely
discussed in the literature. In paper \cite{DGP} it was argued
that matter on the brane can induce a brane-localized curvature
term via quantum corrections, which appears in the low-energy
effective action. An attractive feature of this model is a
modification of gravity at ultra-large scales, which can be very
interesting from the cosmological point of view. Unfortunately a
strong coupling effect was found in the model \cite{Luty,Rubakov}.
Nevertheless, the DGP-proposal can be utilized not only in the
models with infinite extra dimension. For example, in
\cite{Luty,DHR,DGKN} some models with compact extra dimension and
brane-localized curvature terms were discussed. The model proposed
in \cite{DGKN} possesses some very interesting feature - an extra
dimension in this model can be astronomically large. In the
present paper we discuss another model with astronomically large
extra dimension. It is based on the Randall-Sundrum solution for
the metric \cite{RS1} and was discussed in \cite{NPBS} in the case
of sub-millimeter extra dimension. Here we consider this model
from another point of view - in the case of a very large size of
extra dimension.

\section{The setup}
The action of the model considered in \cite{NPBS} has the
following form
\begin{equation}\label{actionRS}
 S = S_g + S_1 + S_2 + S_{ind},
\end{equation}
where $S_g$, $S_1$, $S_2$ and $S_{ind}$ are given by
\begin{eqnarray}\label{actionsRS}
S_g&=& \frac{1}{16 \pi \hat G} \int \left(\hat
R-\Lambda\right)\sqrt{-g}\, d^{4}x dy,\\ \nonumber
 S_1&=&  -\frac{1}{k16 \pi \hat G}\int\left(\tilde R- \Lambda\right)
\delta(y)\sqrt{-\tilde g} d^{4}x dy,\\ \nonumber
 S_2&=&  \frac{1}{k16 \pi \hat G}\int\left(\tilde R-\Lambda\right)
\delta(y-R)\sqrt{-\tilde g} d^{4}x dy,\\ \nonumber S_{ind}&=&
\frac{\Omega^{2}_{ind}}{k16\pi\hat G}\int\tilde R\,
\delta(y)\sqrt{-\tilde g} d^{4}x dy,
\end{eqnarray}
$\Omega_{ind}$ is a dimensionless parameter and
$\Lambda=-12k^{2}$. {Here $\tilde g_{\mu\nu}$ is the induced
metric on the branes and the subscripts 1 and 2 label the branes.}
The model possesses usual $Z_{2}$ orbifold symmetry. We also note
that the signature of the metric $g_{MN}$ is chosen to be
$(-,+,+,+,+)$. Obviously, the model admits the Randall-Sundrum
solution for the metric
\begin{equation}\label{metricrs}
ds^2=  \gamma_{MN} d{x}^M d{x}^N = \gamma_{\mu\nu} {dx^\mu dx^\nu}
+
  dy^2,
\end{equation}
where $\gamma_{\mu\nu}=e^{2\sigma(y)}\eta_{\mu\nu}$,
$\eta_{\mu\nu}$ is the Minkowski metric and {the function}
$\sigma(y) = -k|y|$ in the interval $-R \leq y \leq R$. The
parameter  $k$ is positive and has the dimension of mass. The
function $\sigma$ has the properties
\begin{equation}\label{sigma}
  \partial_4 \sigma = -k\, sign(y), \quad \partial^2_{4}\sigma
  =-2k(\delta(y) - \delta(y-R)) \equiv  -2k\tilde\delta .
\end{equation}

We denote $\hat \kappa = \sqrt{16 \pi \hat G}$, where $\hat G$ is
the five-dimensional gravitational constant, and parameterize the
metric $g_{MN}$ as
\begin{equation}\label{metricpar}
  g_{MN} = \gamma_{MN} + \hat \kappa h_{MN},
\end{equation}
$h_{MN}$ being the metric fluctuations. In \cite{NPBS} the
linearized equations of motion (for the field $h_{MN}$)
corresponding to the action (\ref{actionRS}) were derived. It was
shown, that in the gauge
\begin{equation}\label{unitgauge}
h_{\mu4} =0, \, h_{44} = h_{44}(x) \equiv \phi (x),
\end{equation}
these {\it linearized} equations possess an additional symmetry
under the transformations
\begin{equation}\label{sym1}
h_{\mu\nu}(x,y)\to
h_{\mu\nu}(x,y)+\sigma\gamma_{\mu\nu}\varphi(x)+\frac{1}{2k^2}\left(\sigma+
\frac{1}{2}\right)\partial_{\mu}\partial_{\nu}\varphi(x),
\end{equation}
\begin{equation}\label{sym2}
\phi(x)\to\phi(x)+\varphi(x),
\end{equation}
which {\it do not} belong to the gauge transformations. The
validity of the gauge (\ref{unitgauge}) was thoroughly checked in
\cite{NPBS,BKSV,SV}. It is evident that with the help of
transformations (\ref{sym1}), (\ref{sym2}) we can impose the
condition $\phi(x)\equiv 0$. In other words it is necessary to
examine the next order of perturbation theory to get the equation
for the radion field. But since the radion field does not interact
directly with matter on the brane, it seems that the linear
approximation is not destroyed by the radion. The key feature of
the model is that the only physically relevant case, in which
tachyons and ghost are absent (and the symmetry (\ref{sym1}),
(\ref{sym2}) is preserved) is when the matter (and the induced
term $S_{ind}$) exists {\it on brane~1 only}. Brane~2 (at the
point $y=R$) can be interpreted as a "naked"\ brane, i.e. a brane
without matter on it (see \cite{NPBS}).

The gravity on the brane in the linear approximation is described
by the field $h_{\mu\nu}$, which satisfies the following effective
equation of motion \cite{NPBS}
\begin{eqnarray}\label{mu-nu-dd}
 & &\frac{1}{2}\left(e^{-2\sigma}\Box
 \left[h_{\mu\nu}-\frac{1}{2}\eta_{\mu\nu}h\right]
+ \partial^2_{4} h_{\mu\nu}\right)- 2k^2h_{\mu\nu}+2k
h_{\mu\nu}\tilde\delta - \\ \nonumber &-&
\frac{1}{2k}\tilde\delta\left(e^{-2\sigma}\Box
\left[h_{\mu\nu}-\frac{1}{2}\eta_{\mu\nu}h\right]\right) +
\frac{\Omega^{2}_{ind}}{2k}\delta(y)\left(e^{-2\sigma}\Box
\left[h_{\mu\nu}-\frac{1}{2}\eta_{\mu\nu}h\right]\right)=\\
\nonumber &=&
-\frac{\hat\kappa}{2}\delta(y)t_{\mu\nu}-\frac{\hat\kappa
k}{6\Omega^{2}_{ind}}\left(1-\frac{\tilde\delta}{k}\right)
\left[\eta_{\mu\nu}-\frac{\partial_{\mu}\partial_{\nu}}{\Box}\right]t,
\end{eqnarray}
$t_{\mu\nu}$ denoting the energy-momentum tensor of matter on the
brane~1 and $t=\eta^{\mu\nu}t_{\mu\nu}$. Using this equation, we
can get the equation for the zero mode of $h_{\mu\nu}$, which has
the form $h^0_{\mu\nu}=\alpha_{\mu\nu}e^{2\sigma}$. Let us
multiply equation (\ref{mu-nu-dd}) by $e^{2\sigma}$ and integrate
it over coordinate $y$. Using the orthonormality conditions, which
have the form \cite{NPBS}
\begin{equation}\label{norm1} \int dy
e^{-2\sigma}\left[1-\frac{1}{k}\delta(y)+\frac{1}{k}\delta(y-R)+
\frac{\Omega^{2}_{ind}}{k}\delta(y)\right]\Psi^{l}\Psi^{n}=\delta_{ln},
\end{equation}
we  get
\begin{equation}\label{massl-modes}
\Box\left(\alpha_{\mu\nu}-\frac{1}{2}\eta_{\mu\nu}\alpha\right)=-\frac{\hat\kappa
k}{\Omega^{2}_{ind}}\,t_{\mu\nu}.
\end{equation}
A fully analogous procedure was made in the case of RS1 model in
\cite{SV}.

In the next section we will estimate the effects produced by the
massive modes. We will not solve equation (\ref{mu-nu-dd}), as it
was made in \cite{SV} for the RS1 model, we will thoroughly
estimate the masses of the modes, their wave functions and
coupling constants to matter on the brane. We would like to remind
that we will use other values of parameters, not that used in
\cite{NPBS}.

\section{Wave functions of the massive modes}
It was shown in \cite{NPBS}, that equation for the wave functions
of the massive modes has the form
\begin{eqnarray}\label{mu-nu-tt1}
& &\frac{1}{2}\left(e^{-2\sigma}\Box
h_{\mu\nu}+\partial^2_{4}h_{\mu\nu}\right)
-2k^2h_{\mu\nu}+2k\tilde\delta h_{\mu\nu}-\frac{1}{2k}\tilde\delta
e^{-2\sigma}\Box h_{\mu\nu} +\\ \nonumber &+&
\frac{\Omega^{2}_{ind}}{2k}\delta(y)e^{-2\sigma}\Box h_{\mu\nu}=0.
\end{eqnarray}
Following the footsteps of \cite{NPBS,BKSV}, we  arrive at the
relations:
\begin{eqnarray}\label{besselZ1}
\Psi^{n}(y)=N_{n}\left(N_{0}\left(t_{n}e^{kR}\right)J_{2}\left(t_{n}e^{-\sigma}\right)-
J_{0}\left(t_{n}e^{kR}\right)N_{2}\left(t_{n}e^{-\sigma}\right)\right),
\end{eqnarray}
where $N_{n}$ is a normalization constant, and
\begin{eqnarray}\label{eigenvalues2}
& &N_{0}\left(t_{n}e^{kR}\right)J_{0}\left(t_{n}\right)-
J_{0}\left(t_{n}e^{kR}\right)N_{0}\left(t_{n}\right)+\\ \nonumber
&+&
\Omega^{2}_{ind}\left[N_{0}\left(t_{n}e^{kR}\right)J_{2}\left(t_{n}\right)-
J_{0}\left(t_{n}e^{kR}\right)N_{2}\left(t_{n}\right)\right] =0.
\end{eqnarray}
Here $h_{\mu\nu}(x,y)=\sum \Psi^{n}(y)h^{n}_{\mu\nu}(x)$,
$\Psi^{n}(y)$ is the wave function of the corresponding massive
mode, $t_{n}=\frac{m_{n}}{k}$ and $m_{n}$ is such that $\Box
h^{n}_{\mu\nu}(x)=m_{n}^{2}h^{n}_{\mu\nu}(x)$. Equation
(\ref{eigenvalues2}) defines the mass spectrum of the theory.

Let us choose $\Omega_{ind}$ such that $\Omega_{ind}>>1$ (we will
specify its value later) and $kR\approx 1$. For the relatively
large values of $\frac{m}{k}$ one can use the expansions for the
Bessel functions for large values of arguments (see, for example,
\cite{Korn}). In this approximation masses of the modes are
defined by
\begin{equation}\label{eigen-t}
\left(e^{kR}-1\right)\frac{m_{n}}{k}\approx \pi n,
\end{equation}
where $n$ is integer. Small corrections to these values are such
that
\begin{equation}\label{psi-t}
\Psi^{n}(0)\sim N_{n}\frac{k^2}{{m_{n}}^2\Omega_{ind}^{2}},
\end{equation}
This formula can be obtained from equation (\ref{eigenvalues2})
(see Appendix~1). It seems that this form of the wave functions
(inverse proportional to $m_{n}^2$) can be obtained only if there
exist brane-localized curvature terms. It is necessary to note
that since the main purpose of \cite{NPBS} was to examine the
structure of linearized equations of motion, only a brief analysis
of $\Psi^{n}(0)$, resulted in a very rough estimate
$\Psi^{n}(0)\sim\frac{1}{\Omega_{ind}}$, was carried out in that
paper. This estimate is sufficient for not rejecting the model in
the case of sub-millimeter extra dimension. Equation (\ref{psi-t})
shows that contributions of the massive modes appear to be
suppressed much stronger, and for the case of sub-millimeter extra
dimension (an example in \cite{NPBS}) an effective theory on the
brane actually describes massless 4-D gravity. At the same time
the form of (\ref{psi-t}) is very important for the case of very
large extra dimension, which will be considered below.

It is not difficult to show that integral in (\ref{norm1}) is
always positive and $N_{n}\sim\frac{m_{n}}{\sqrt{k}}$ for
relatively large $m_{n}$ (see Appendix~2). Thus the form of
interaction of the massive modes with matter on the brane looks
like
\begin{equation}\label{interaction6}
\frac{1}{2}\int_{brane}d^4x\left(\hat\kappa\sum_{n\le
j}\Psi^{n}(0)h^n_{\mu\nu}(x)t^{\mu\nu}+\frac{1}{M_{Pl}
\Omega_{ind}}\sum_{n>j}\frac{a_{n}}{n}h^n_{\mu\nu}(x)t^{\mu\nu}\right),
\end{equation}
where $a_{n}\sim 1$, $M_{Pl}=\Omega_{ind}/\sqrt{\hat G k}$ and $j$
is such that for $n>j$ one can use expansion formulas for the
Bessel functions with good accuracy. One can see, that coupling
constants of the massive modes are suppressed by the factor
$\Omega_{ind}$ and strongly depend on the number of a mode. It is
not difficult to show that for $n\le j$ the factor
$\hat\kappa\Psi^{n}(0)$ is of the order of
$1/(\Omega_{ind}M_{Pl})$. This estimate can be obtained from
(\ref{norm1}), (\ref{besselZ1}) and (\ref{eigenvalues2}) using the
fact that the Bessel and Neumann functions are of the order of
unity or lesser for the masses, which are the solutions of
(\ref{eigenvalues2}); and the fact that combination in the square
brackets in (\ref{eigenvalues2}) is equal to $\Psi^{n}(0)$
(accurate to the factor $N_{n}$).

Now let us discuss the possible values of parameters of the model.
We would like to mention, that effective theory exactly of this
type appears in the model considered in \cite{DGKN}, so we will
use some results obtained there. Because of the fact that coupling
of Kaluza-Klein modes to matter on the brane is defined by the
factors $\left(\Psi^{n}(0)\right)^2$ and the sum $\sum
\frac{1}{n^2}$ is rapidly convergent, one may not to worry about
the strong affection of massive modes because of their quantity
(it was thoroughly checked in \cite{DGKN}). This shows an
importance of the fact that (\ref{psi-t}) is inverse proportional
to $m_{n}^2$ and $N_{n}$ is proportional to $m_{n}$. Let us
consider $R$ of the order, say, of the size of the Sun. For this
case constraints on $\Omega_{ind}$ following from the
astrophysical experiments are very severe (see \cite{DGKN}) -
$\Omega_{ind}>10^{5}$ (in our notations). The multi-dimensional
Planck mass $M$ should be of the order of $10\, GeV$ to get the
proper four-dimensional gravitational constant, which is defined
by (\ref{massl-modes}). Thus the hierarchy problem is solved in an
analogous way to that proposed in \cite{ADD,Ant}. In this case the
massive modes can become apparent on the distances of the order of
extra dimension's size. It can lead to some slight affection on
the Mercury motion.

Of course, the size of extra dimension can be much lesser: for
example, from the size of the Earth down to the sub-millimeter
scales and even smaller. In this case constraints on the value of
$\Omega_{ind}$ following from the astrophysical data can be
substantially relaxed, and the only claims are that the values of
$\hat G$ and $\Omega_{ind}$ with given $R$ should give the proper
value of four-dimensional gravitational constant and correspond to
the constraints obtained from the top-table gravitational
experiments. Because of suppression of the massive modes, the
phenomenology of this model differs considerably from that one of
any standard "large extra dimensions" scenario (this issue was
discussed in detail in \cite{DGKN}).

\section{Discussion and final remarks}
In this paper we constructed a self-consistent model with
astronomically large compact extra dimension. One may ask: why we
use warped geometry, whereas the same effect can be achieved in
the model with flat background \cite{DGKN}? Firstly, in the paper
\cite{DGKN} the radion field ($44$-component of the metric
fluctuations) was taken into account not throughout all of
calculations. It is well-known that existence of this field can
change the theory considerably, as it was in the case of DGP-model
\cite{Luty,Rubakov}. It is not probable that the same problem can
arise in the model described in \cite{DGKN}, nevertheless it is
necessary to solve corresponding equations of motions with regard
to the radion field more accurately, checking the validity of
imposed gauge conditions. In our case considered above the radion
field is absent in the linear approximation and one can forget
about it (and about the problems with ghosts, see
\cite{Luty,DubLib,Padilla}) at least at the classical level.
Secondly, the main advantage of warped geometry is that it ensures
the proper tensor structure of the massless graviton propagator
(in the case of flat geometry this property seems to be forbidden
by the equation for the $44$-component of metric fluctuations). It
is also necessary to note, that the value of $\Omega_{ind}$ was
chosen to correspond an appropriate value in \cite{DGKN}. More
realistic minimal value, corresponding to the experimental data,
can be a little bit larger (because of the warped background, see
(\ref{interaction6})). But it is evident that this difference
should not be considerable. Some other interesting features of
this setup were mentioned in \cite{NPBS}.

Now let us discuss our choice for the size of extra dimension.
Solution of equations of motion for the $\mu\nu$-component of
metric fluctuations includes some term (see \cite{NPBS}), which is
a pure gauge from the four-dimensional point of view of an
observer on the brane. This term is proportional to the factor
$\frac{1}{k^2}$, and in the case of very large size of extra
dimension (and correspondingly very small $k$) the linear
approximation brakes down. Exactly the same thing happens in
massive gravity at the small graviton mass limit, see for example
\cite{massive}. Thus the size of extra dimension of the order of
the Sun's size seems to be "safe"\ from this point of view. For a
cases of much larger size of extra dimension it is necessary to
carry out a more detailed analysis including examination of the
non-linear effects.

We would like to mention, that only the case $e^{kR}\sim 1$ was
considered in this paper. The case $e^{kR}>>1$ seems to be quite
interesting, because there can appear an additional suppression of
contributions of the massive modes by the exponential factor,
which could relax the constraints on $\Omega_{ind}$. At the same
time the lowest modes will have lower masses $\sim ke^{-kR}$ in
this case. Anyway this situation calls for more detailed
investigation. Also it would be interesting whether the symmetry
analogous to (\ref{sym1}), (\ref{sym2}) could arise in other
models, for example, with infinite extra dimension, which admit
modification of gravity at ultra-large scales.

\bigskip
{\large \bf Acknowledgments}
\medskip \\
The author is grateful to I.P.~Volobuev for valuable discussions.
The work was supported by the RFBR grant 04-02-16476, by the grant
UR.02.02.503 of the scientific program "Universities of Russia",
and by the grant NS.1685.2003.2 of the Russian Federal Agency for
Science.

\section{Appendix~1}
Let us calculate estimates (\ref{eigen-t}) and (\ref{psi-t}) for
the case $\frac{m}{k}>>1$ and $kR\approx 1$. For large values of
arguments we can use the following expansions for the Bessel
functions \cite{Korn}
\begin{equation}\label{Bes1}
J_{0}(t)\approx\sqrt{\frac{2}{\pi
t}}\left[\cos\left({t-\frac{\pi}{4}}\right)+\frac{1}{8t}\sin\left({t-\frac{\pi}{4}}\right)\right],
\end{equation}
\begin{equation}\label{Bes2}
N_{0}(t)\approx\sqrt{\frac{2}{\pi
t}}\left[\sin\left({t-\frac{\pi}{4}}\right)-\frac{1}{8t}\cos\left({t-\frac{\pi}{4}}\right)\right],
\end{equation}
\begin{equation}\label{Bes3}
J_{2}(t)\approx\sqrt{\frac{2}{\pi
t}}\left[\cos\left({t-\frac{\pi}{4}-\pi}\right)-\frac{15}{8t}\sin\left({t-\frac{\pi}{4}-\pi}\right)\right],
\end{equation}
\begin{equation}\label{Bes4}
N_{2}(t)\approx\sqrt{\frac{2}{\pi
t}}\left[\sin\left({t-\frac{\pi}{4}-\pi}\right)+\frac{15}{8t}\cos\left({t-\frac{\pi}{4}-\pi}\right)\right].
\end{equation}
Substituting (\ref{Bes1})-(\ref{Bes4}) into (\ref{eigenvalues2}),
we get
\begin{eqnarray}\nonumber
\sin\left((\lambda
-1)t\right)+\left[\frac{1}{8t}-\frac{1}{8\lambda
t}\right]\cos\left((\lambda -1)t\right)=\\ \nonumber =-
\Omega_{ind}^2\left(\sin\left((\lambda
-1)t\right)-\left[\frac{15}{8t}+\frac{1}{8\lambda
t}\right]\cos\left((\lambda -1)t\right)\right),
\end{eqnarray}
where $t=\frac{m}{k}$, $\lambda=e^{kR}$. For the relatively large
$t$ solutions of this equation in the zero approximation are
defined by $$ t_{n}\approx \frac{\pi n}{\lambda-1}, $$ where $n$
is integer. Corrections $\Delta t_{n}$ to $t_{n}$ at the first
order are defined by the equation $$ (-1)^n\Delta
t_{n}\left(\lambda-1\right)\left(\Omega_{ind}^2+1\right)=(-1)^n\left[
-\frac{1}{8t_{n}}+\frac{1}{8\lambda
t_{n}}+\frac{\Omega_{ind}^2}{8\lambda
t_{n}}+\frac{15\Omega_{ind}^2}{8 t_{n}}\right], $$ from which one
gets $$ \Delta t_{n}\approx \frac{1}{(\lambda
-1)t_{n}}\left[\frac{15}{8}+\frac{1}{8\lambda}-\frac{16}{8\Omega_{ind}^2}\right].
$$ Finally for the function $\Psi^{n}(0)$ we have $$
\Psi^{n}(0)\approx N_{n}\sqrt{\frac{2}{\pi t_{n}}
}\sqrt{\frac{2}{\pi t_{n}\lambda}}\left(\frac{16}{8\Omega_{ind}^2
t_{n}}\right)(-1)^n=\frac{4N_{n}}{\pi
t_{n}^2\Omega_{ind}^2\sqrt{\lambda}}(-1)^n. $$

\section{Appendix~2}
Let us show, that
\begin{equation}\label{Bes5}
\int dy
e^{-2\sigma}\left[1-\frac{1}{k}\delta(y)+\frac{1}{k}\delta(y-R)+
\frac{\Omega^{2}_{ind}}{k}\delta(y)\right]\left(\Psi\right)^{2}>0,
\end{equation}
if $\Psi$ satisfies the equation
\begin{eqnarray}\label{Bes6}
\frac{1}{2}\left(e^{-2\sigma}m^{2}\Psi+\partial^2_{4}\Psi\right)
-2k^2\Psi+2k\tilde\delta \Psi-\\
\nonumber-\frac{1}{2k}\tilde\delta e^{-2\sigma}m^{2}\Psi +
\frac{\Omega^{2}_{ind}}{2k}\delta(y)e^{-2\sigma}m^{2}\Psi=0.
\end{eqnarray}
Here
\begin{eqnarray}\nonumber
\Psi(y)&=&NZ_{2}(y),\\ \nonumber
Z_{2}(y)&=&N_{0}\left(\frac{m}{k}e^{kR}\right)J_{2}\left(\frac{m}{k}e^{-\sigma}\right)-
J_{0}\left(\frac{m}{k}e^{kR}\right)N_{2}\left(\frac{m}{k}e^{-\sigma}\right),
\end{eqnarray}
$N$ is normalization constant. It is easy to show that
\begin{eqnarray}\label{Bes7}
\int_{-R}^{R} dy
e^{-2\sigma}Z_{2}^{2}(y)=\left[t=\frac{m}{k}e^{-\sigma}\right]=\frac{2k}{m^{2}}\int_{t_0}^{t_R}tZ_{2}^{2}(t)dt,
\end{eqnarray}
where $t_0=\frac{m}{k}$, $t_R=\frac{m}{k}e^{kR}$. With the help of
the Bessel equation for the function $Z_{2}(t)$ $$ \ddot
Z_{2}(t)+\frac{\dot
Z_{2}(t)}{t}+\left(1-\frac{4}{t^{2}}\right)Z_{2}(t)=0, $$ где
$\dot Z_{2}(t)=\frac{d}{dt}Z_{2}(t)$, one can get for the integral
(\ref{Bes7})
\begin{eqnarray}\nonumber
\frac{2k}{m^{2}}\int_{t_0}^{t_R}Z_{2}(t)\left[\frac{4Z_{2}(t)}{t}-\dot
Z_{2}(t)-t\ddot Z_{2}(t)\right]dt=\\ \nonumber=
\frac{2k}{m^{2}}\int_{t_0}^{t_R}\left[t\dot
Z_{2}^2(t)+\frac{4Z_{2}^2(t)}{t}\right]dt+\left.\frac{2k}{m^{2}}\left(-t\dot
Z_{2}(t)Z_{2}(t)\right)\right|_{t_{0}}^{t_{R}},
\end{eqnarray}
where integration by parts were made. Using the boundary
conditions for $Z_{2}(t)$, which follow from the equation
(\ref{Bes6}) and have the form
\begin{eqnarray}\nonumber
\left. t\dot
Z_{2}(t)+2Z_{2}(t)-\frac{t^2}{2}Z_{2}(t)\right|_{t=t_{R}}=0,\\
\nonumber \left. t\dot
Z_{2}(t)+2Z_{2}(t)-\frac{t^2}{2}Z_{2}(t)+\Omega_{ind}^{2}\frac{t^2}{2}Z_{2}(t)\right|_{t=t_{0}}=0,
\end{eqnarray}
we can show that integral (\ref{Bes5}) is equal to (accurate to
the factor $N^2$)
\begin{equation}\label{Bes8}
\frac{2k}{m^{2}}\left(\int_{t_0}^{t_R}\left[t\dot
Z_{2}^2(t)+\frac{4Z_{2}^2(t)}{t}\right]dt+\left.2\left[Z_{2}^2(t)\right]\right|_{t_{0}}^{t_{R}}\right).
\end{equation}
Substituting the recurrent formula for the Bessel function
\begin{eqnarray}\nonumber
\dot Z_{2}(t)=Z_{1}(t)-\frac{2}{t}Z_{2}(t),
\end{eqnarray}
where
\begin{eqnarray}\nonumber
Z_{1}(t)&=&N_{0}\left(\frac{m}{k}e^{kR}\right)J_{1}\left(t\right)-
J_{0}\left(\frac{m}{k}e^{kR}\right)N_{1}\left(t\right)
\end{eqnarray}
into (\ref{Bes8}), we get
\begin{equation}\label{Bes9}
\frac{2k}{m^{2}}\left(\int_{t_0}^{t_R}\left[tZ_{1}^2(t)-4\dot
Z_{2}(t)Z_{2}(t)\right]dt+\left.2\left[Z_{2}^2(t)\right]\right|_{t_{0}}^{t_{R}}\right)
=\frac{2k}{m^{2}}\int_{t_0}^{t_R}tZ_{1}^2(t)dt.
\end{equation}
It is evident that this integral is always positive. Thus the
integral in (\ref{Bes5}) is also positive for any real $N$.

Now let us calculate (\ref{Bes9}) for the relatively large $m$.
Using the formulas analogous to (\ref{Bes1})-(\ref{Bes4}) (see
\cite{Korn}) one can easily show that
$$Z_{1}(t)\approx\frac{2}{\pi\sqrt{t_{R}}\sqrt{t}}\cos(t-t_R).$$
Thus
$$\frac{2k}{m^{2}}\int_{t_0}^{t_R}tZ_{1}^2(t)dt\approx\frac{4k}{m^{2}\pi^2}
\frac{t_R-t_0}{t_R}=\frac{4k}{m^{2}\pi^2}(1-e^{-2kR})$$ and
$N\sim\frac{m}{\sqrt{k}}$ for $kR\approx 1$.


\begin{thebibliography}{99}
\bibitem{DGP}
G.~R.~Dvali, G.~Gabadadze, M.~Porrati, Phys.\ Lett. B485 (2000)
208 [arXiv:hep-th/0005016].

\bibitem{Luty}
M.~A.~Luty, M.~Porrati, R.~Rattazzi, JHEP 0309 (2003) 029
[arXiv:hep-th/0303116].

\bibitem{Rubakov}
V.A.~Rubakov, "Strong coupling in brane-induced gravity in five
dimensions", arXiv:hep-th/0303125.

\bibitem{DHR}
H.~Davoudiasl, J.~L.~Hewett, T.~G.~Rizzo, JHEP 0308 (2003) 034
[arXiv:hep-ph/0305086].

\bibitem{DGKN}
G.R.~Dvali, G.~Gabadadze, M.~Kolanovic, F.~Nitti, Phys. Rev. D64
(2001) 084004 [arXiv:hep-ph/0102216].

\bibitem{RS1}
L.~Randall, R.~Sundrum, Phys. Rev. Lett. 83 (1999) 3370
[arXiv:hep-ph/9905221].

\bibitem{NPBS}
M.N.~Smolyakov, Nucl. Phys. B695 (2004) 301
[arXiv:hep-th/0403034].

\bibitem{BKSV}
E.E.~Boos, Yu.A.~Kubyshin, M.N.~Smolyakov, I.P.~Volobuev, Theor.
Math. Phys. 131(2) (2002) 629 [arXiv:hep-th/0105304].

\bibitem{SV}
I.P.~Volobuev, M.N.~Smolyakov, Theor. Math. Phys. 139(1) (2004)
458 [arXiv:hep-th/0208025].

\bibitem{Korn}
G.~Korn, T.~Korn, "Mathematical handbook for scientists and
engeneers", (McGraw-Hill, 1968).

\bibitem{ADD} N.~Arkani-Hamed, S.~Dimopoulos, G.~Dvali, Phys. Lett.
B429 (1998) 263 [arXiv:hep-ph/9803315].

\bibitem{Ant}
I.~Antoniadis, N.~Arkani-Hamed, S.~Dimopoulos, G.R.~Dvali, Phys.
Lett. B436 (1998) 257 [arXiv:hep-ph/9804398].

\bibitem{DubLib}
S.~L.~Dubovsky, M.~V.~Libanov, JHEP 0311 (2003) 038
[arXiv:hep-th/0309131].

\bibitem{Padilla}
A.~Padilla, Class. Quant. Grav. 21 (2004) 2899
[arXiv:hep-th/0402079].

\bibitem{massive}
M.~Porrati, "Massive gravity in ADS and Minkowski backgrounds",
arXiv:hep-th/0409172.

\end{thebibliography}
\end{document}